\documentclass[aps,10pt,pra,twocolumn,showpacs,amsmath,amssymb, superscriptaddress]{revtex4-1}

\usepackage{graphicx}
\usepackage{color}
\usepackage{transparent}
\usepackage{hyperref}
\usepackage{txfonts}
\usepackage{amsmath}
\usepackage{braket}

\begin{document}

\title{Synchronization of an optomechanical system to an external drive}

\author{Ehud Amitai}
\affiliation{Department of Physics, University of Basel, Klingelbergstrasse 82, 4056 Basel, Switzerland}
\author{Niels L\"{o}rch}
\affiliation{Department of Physics, University of Basel, Klingelbergstrasse 82, 4056 Basel, Switzerland}
\author{Andreas Nunnenkamp}
\affiliation{Cavendish Laboratory, University of Cambridge, Cambridge CB3 0HE, United Kingdom}
\author{Stefan Walter}
\affiliation{Max Planck Institute for the Science of Light, Staudtstrasse 2, 91058 Erlangen, Germany}
\author{Christoph Bruder}
\affiliation{Department of Physics, University of Basel, Klingelbergstrasse 82, 4056 Basel, Switzerland}

\begin{abstract}
Optomechanical systems driven by an effective blue detuned laser can exhibit self-sustained oscillations of the mechanical oscillator. These self-oscillations are a prerequisite for the observation of synchronization. Here, we study the synchronization of the mechanical oscillations to an external reference drive. We study two cases of reference drives: (1) An additional laser applied to the optical cavity; (2) A mechanical drive applied directly to the mechanical oscillator. Starting from a master equation description, we derive a microscopic Adler equation for both cases, valid in the classical regime in which the quantum shot noise of the mechanical self-oscillator does not play a role. Furthermore, we numerically show that, in both cases, synchronization arises also in the quantum regime. The optomechanical system is therefore a good candidate for the study of quantum synchronization.
\end{abstract}
\pacs{05.45.Xt, 42.65.-k, 07.10.Cm}
\maketitle

\section{Introduction}

Synchronization is the phenomenon in which a limit-cycle oscillator, i.e.,~an oscillator with fixed amplitude and free phase, develops a phase preference when weakly coupled to a drive or to other self-oscillating systems~\cite{Pikovsky, Balanov}. This phenomenon is prevalent in all the natural sciences, manifesting itself in, for example, adjustment of the circadian rhythm in many living systems or fireflies blinking in unison~\cite{Strogatz}. 

In recent years there has been considerable interest in the topic of quantum synchronization~\cite{Walter1, Walter2, Lee1, Lee2, Loerch, Weiss2, Bastidas, Ameri, Lorch3, Davis, Xu, Roth, Zhu, Hush, Weiss, Ludwig, Yin, Gian}, i.e.,~the synchronization of self-oscillators operating in the quantum regime. There has been extensive research done on the paradigmatic example of a van der Pol oscillator~\cite{Walter1, Walter2, Lee1, Lee2, Loerch, Weiss2, Bastidas, Ameri, Lorch3}. 
Other platforms have been used to study quantum synchronization as well, among which are micromasers~\cite{Davis}, atomic ensembles~\cite{Xu, Roth}, interacting quantum dipoles~\cite{Zhu}, trapped ions~\cite{Lee1, Hush}, and optomechanical systems~\cite{Weiss, Ludwig, Yin}.

In an optomechanical system electromagnetic cavity modes are coupled to mechanical motion. In its most basic setup, an optomechanical system is made of a single laser-driven cavity mode which couples to a single mechanical mode via, e.g.,~radiation pressure~\cite{Aspelmeyer}. The dynamics of the system crucially depends on the frequency of the laser driving the cavity. A laser field tuned to the red side of the cavity frequency is used for back-action cooling as well as for state transfer~\cite{Teufel, Chan, Palomaki}, while resonant driving is used, e.g.,~for position sensing~\cite{Purdy}. When blue detuned, the laser drive can induce the parametric instability, leading in turn to self-sustained oscillations of the mechanical oscillator. These self-sustained oscillations have been studied in both the classical and the quantum regimes~\cite{Marquardt, Metzger, Qian, Loerch2, Rodrigues, Armour}. For that reason, the optomechanical system may exhibit synchronization when coupled to an external drive (an additional external drive, in contrast to the laser driving the self-oscillations), to another optomechanical system, or as part of an array of optomechanical systems, as was theoretically shown in the classical regime~\cite{Heinrich, Holmes}. Synchronization of an optomechanical system to an external drive~\cite{Shlomi}, of two optomechanical systems~\cite{Zhang1} and even of small arrays of up to seven such systems~\cite{Zhang2} have been demonstrated experimentally. In the quantum regime the synchronization of two optomechanical systems has been studied theoretically~\cite{Weiss}, as well as the synchronization of an array of such systems~\cite{Ludwig} within a mean-field approach was used.

In this work, we theoretically study the synchronization of the mechanical self-oscillator to an external reference drive. We examine two different reference drives: (1) An additional laser applied to the optical cavity. Under an appropriate rotating-wave approximation, this may also be implemented by modulating the power of the laser inducing the mechanical self-oscillations, as was experimentally done in Ref.~\cite{Shlomi};
(2) A mechanical drive applied directly onto the mechanical oscillator, which could for instance be realized with a piezoelectric element attached to the mechanical oscillator. 

For both cases, our starting point of the analysis is the microscopic master equation. We then use the laser theory for optomechanical limit cycles~\cite{Loerch2} to derive an equation of motion (EOM) for the phase distribution of the mechanical self-oscillator. We show that in both cases, in a relevant parameter regime, the Adler equation emerges from the EOM. The Adler equation is a differential equation for the phase difference between the self-oscillator and the reference drive, known to describe synchronization. For the optical reference drive, this is the first time a microscopically derived Adler equation is discussed. For the mechanical reference drive, it reproduces a result in Ref.~\cite{Heinrich}. We then continue to show numerically, for both cases, that in the quantum parameter regime an ``Arnold tongue" exists, a standard signature of synchronization~\cite{Pikovsky, Balanov}. This suggests the optomechanical system is a good candidate for the study of synchronization in the quantum regime.

This manuscript is organized as follows: We describe the system under investigation, composed of an optomechanical system and an additional reference drive in Sec.~\ref{Sec:System}. Section~\ref{Sub:Analytical_calculation} contains the analytical derivation of the microscopic Adler equations. A major part of this derivation is done by applying the laser theory for optomechanical limit cycles~\cite{Loerch2} to this problem. This is presented in the Appendix. In Sec.~\ref{quant_sync} we demonstrate numerically that synchronization is expected also in a quantum parameter regime. 

\section{The system} \label{Sec:System}

\begin{figure}[t]
\includegraphics[width=\columnwidth]{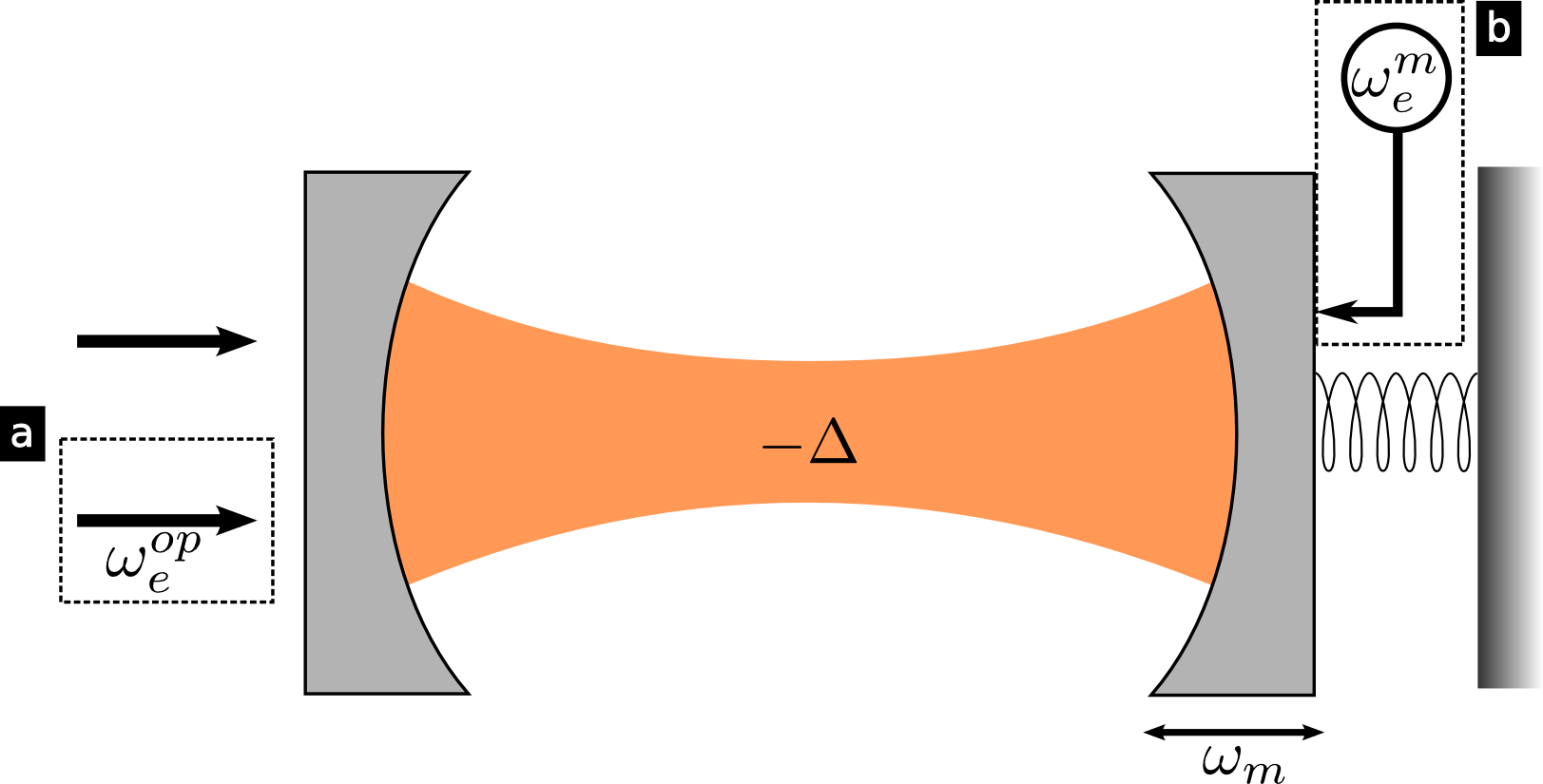}\caption{Schematics of a generic optomechanical system. In a rotating frame with frequency $\omega_L$, obtained by applying the unitary transformation $\hat{U}=\exp \left(i\omega_L t a^\dagger a\right)$, the cavity with frequency $-\Delta$ is driven by a time-independent laser, depicted by the black arrow to the left of the cavity. In the self-oscillatory regime of the mechanical oscillator with natural frequency $\omega_m$, the mechanical oscillator may synchronize to an additional optical drive with frequency $\omega_e^{op}$ as depicted in dashed box (a), or to a mechanical drive with frequency $\omega_e^m$ as depicted in dashed box (b). Note that $\omega_e^{op}$ is given in the rotating frame, while the application of $\hat{U}$ leaves both $\omega_e^m$ and $\omega_m$ identical in both frames. 
\label{Fig:Optomechanical_System}}
\end{figure}

The standard Hamiltonian of an optomechanical system in which the position of the mechanical oscillator parametrically modulates the frequency of an electromagnetic cavity is given in a frame rotating with the frequency of the laser drive, $\omega_L$, by~\cite{Aspelmeyer}
\begin{equation} \label{Bare_Hamiltonian}
	H
=
	\omega_m b^\dagger b
-
	\Delta a^\dagger a
-
	g_0 a^\dagger a 
	\left(
		b
+
		b^\dagger
	\right)
-
	iA_L
	\left(
		a
	-
		a^\dagger
	\right),
\end{equation}
where $a^\dagger$ and $a$ are the creation and annihilation operators of photons in the cavity, $b^\dagger$ and $b$ are the creation and annihilation operators of phonons in the mechanical resonator, $\omega_m$ is the mechanical frequency of oscillation, $\Delta=\omega_L-\omega_c$ is the detuning from cavity resonance at $\omega_c$ of the driving field with strength $A_L$, $g_0$ is the single photon coupling constant, and we have set $\hbar=1$. A schematic figure of the system is shown in Fig.~\ref{Fig:Optomechanical_System}. The frame rotating with laser drive $\omega_L$ is obtained by applying the unitary transformation $\hat{U}=\exp \left(i\omega_L t a^\dagger a\right)$, which generates the Hamiltonian $\hat{U}H_\text{old}\hat{U}^\dagger-i\hat{U}\partial\hat{U}^\dagger/\partial t$.

The dissipation of the two oscillators (the mechanical resonator and the optical cavity) can be described via the master equation,
\begin{equation}
	\frac{\mathrm{d} \rho}{\mathrm{d} t}
=
-
	i
	\left[
		H,\rho
	\right]
+
	\mathcal{L}_m
	\rho
+
	\mathcal{L}_c
	\rho,
			\label{temp_master1}
\end{equation}
with the Lindblad operators
\begin{equation}
	\mathcal{L}_m\rho
=
	\gamma_m
	(n_\text{th}+1)
	\mathcal{D}[b]
	\rho
+
	\gamma_m
	n_\text{th}
	\mathcal{D}
	[b^\dagger]
	\rho,
\end{equation}
and
\begin{equation}
	\mathcal{L}_c \rho
=
	\gamma_c
	\mathcal{D}[a]
	\rho,
\end{equation}
where $\gamma_m$  and $\gamma_c$ are the amplitude damping rates of the mechanical oscillator and the electromagnetic cavity correspondingly, $n_\text{th}$ is the mean phonon number in thermal equilibrium, and $D[x]\rho=x\rho x^\dagger -\left( \rho x^\dagger x +x^\dagger x \rho\right)/2$.

In this work we would like to study the synchronization of the mechanical element of the optomechanical system to an external drive. 
We consider two cases:


\textit{Case (1) - Optical laser drive} - We introduce an additional optical laser reference field, with frequency $\omega_e^{op}$ given in a frame rotating with frequency $\omega_L$, and strength $A_e^{op}$, by adding the term

\begin{equation} \label{optical_drive}
	H^\text{op}
=
-
	iA_e^{op}
	\left(
		e^{i\omega_e^{op} t}
		a
	-
		e^{-i\omega_e^{op} t}
		a^\dagger
	\right)
\end{equation}
to the Hamiltonian appearing in Eq.~(\ref{temp_master1}). This is depicted in dashed box (a) in Fig.~\ref{Fig:Optomechanical_System}. This Hamiltonian can be realized by an additional optical laser, or, if the mechanical frequency $\omega_m$ is large enough such that a rotating-wave approximation can be used, by periodically modulating the power of the optical laser causing the mechanical self-oscillations, as seen in Ref.~\cite{Shlomi}.


\textit{Case (2) - Mechanical drive} - A mechanical reference drive with frequency $\omega_e^m$ and strength $A_e^m$ can be applied directly onto the mechanical oscillator, e.g.,~by introducing a piezoelectric component as depicted in dashed box (b) in Fig.~\ref{Fig:Optomechanical_System}. In analogy to case (1), we add the term 
\begin{equation} \label{mechanical_drive}
	H^\text{m}
=
-
	iA_e^m
	\left(
		e^{i\omega_e^m t}
		b
	-
		e^{-i\omega_e^m t}
		b^\dagger
	\right)
\end{equation}
to the Hamiltonian appearing in Eq.~(\ref{temp_master1}). 


\begin{figure}[t]
\includegraphics[width=\columnwidth]{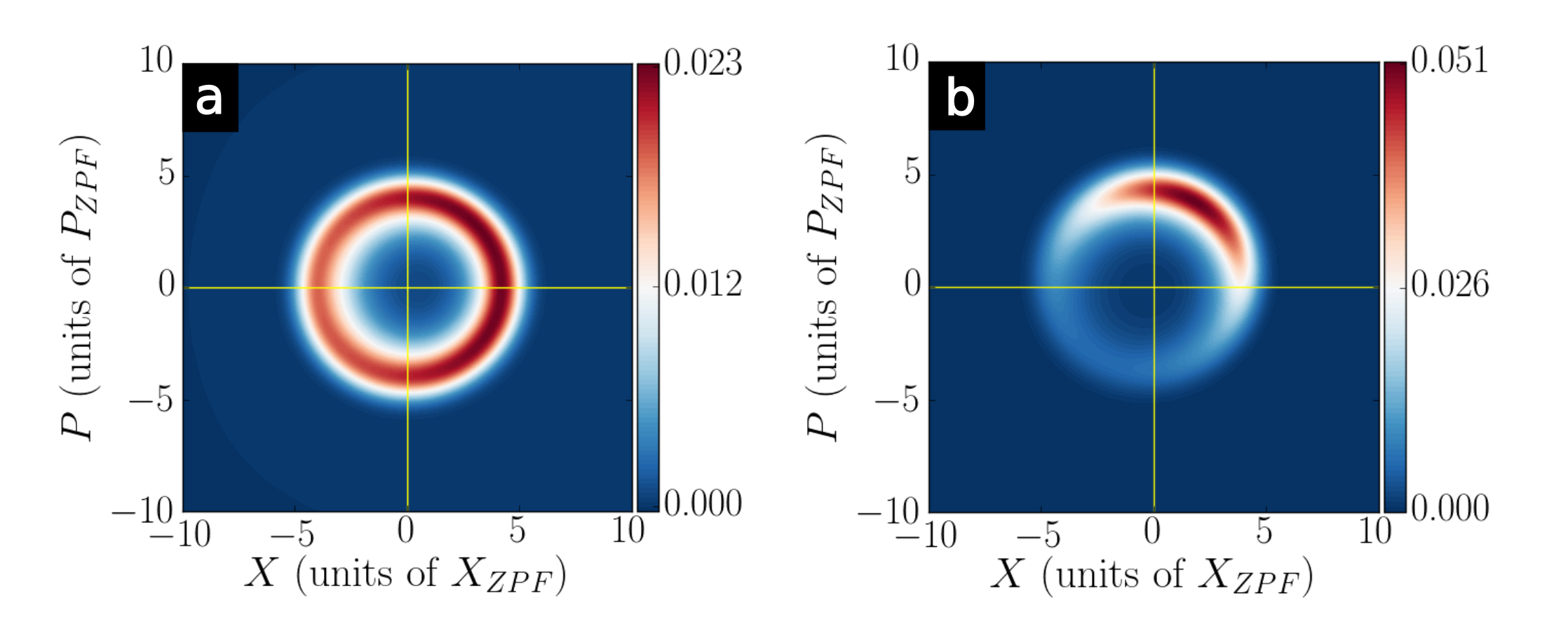}\caption{Wigner function representation of (a) self-sustained oscillations in the mechanical oscillator and of (b) tendency towards phase-locking of the mechanical oscillator to the phase of a synchronizing reference drive. The parameters used for both plots are $(g_0, \gamma_c, \gamma_m, A_L, \Delta, n_\text{th})=(0.3, 0.3, 0.015, 0.4, 0, 0)\times \omega_m$, where the parameters of the external optical drive in (b) are $(A_e^{op}, \omega_e^{op})=(0.08, 0.98)\times \omega_m$. \label{Fig:Wigner_Bare}}
\end{figure}

\textit{Self-oscillations in the optomechanical system.-} An optomechanical system driven by an effective blue-detuned laser may give rise to self-sustained oscillations in the mechanical oscillator~\cite{Marquardt, Metzger, Qian, Loerch2, Rodrigues, Armour}. These self-oscillations are a prerequisite for studying synchronization. They can be illustrated in phase-space via the Wigner function phase-space distribution. A Wigner function representation for a specific self-sustained oscillation in the mechanical oscillator is shown in Fig.~\ref{Fig:Wigner_Bare}~(a). 

Under the influence of a reference drive, the mechanical self-oscillator may develop a phase-preference as it tends towards locking onto the phase of the drive. For an additional optical laser drive, as in case~(1), the Wigner representation for a mechanical self-oscillator showing such phase-preference is shown in Fig.~\ref{Fig:Wigner_Bare}~(b)\footnote{The influence of the reference drive, ideally, should not change the amplitude of the self-oscillator. In practice however, there is some influence. In this work we make sure that the reference drive does not change the amplitude of the unsynchronized limit cycle by more than 10\%}.

\section{Synchronization - Analytical Calculation}    \label{Sub:Analytical_calculation}
In the following section, it is our goal to derive an analytical description for the synchronization of the mechanical self-oscillator to a reference drive. To do so, we apply the laser theory for optomechanical limit cycles \cite{Loerch2} to the current problem, Eq.~(\ref{temp_master1}), in which an additional reference drive, Eq.~(\ref{optical_drive}) or Eq.~(\ref{mechanical_drive}), is influencing the optomechanical limit cycle. This approach is based on the assumption that the dynamics of the cavity adiabatically follows the dynamics of the mechanical oscillator. It allows us to obtain an equation of motion (EOM) for the phase distribution of the self-oscillator, $\sigma(r,\phi)$, where $r$ and $\phi$ are the mechanical phase-space variables representing the radius and the phase of the self-oscillator. To keep this manuscript coherent and focused on synchronization, we shift the derivation of the relevant EOMs to the Appendix. Here in the main text, we will use the derived EOMs as a starting point.

\textit{Case (1) - Optical laser drive} - As explained in the Appendix, the EOM for $\sigma(r,\phi)$ is valid when the dynamics of the cavity adiabatically follows the dynamics of the mechanical oscillator, the optomechanical coupling is small, $g_0\ll\omega_m$, and the thermal- and quantum shot-noise does not play a role. In a rotating frame with frequency $\omega_m$, the EOM for $\sigma(r,\phi)$ is of the form
\begin{equation}  \label{FP_equation_main}
	\dot{\sigma}
=
-\left(
	\partial_r \mu_r
+
	\partial_\phi \mu_\phi
\right)
\sigma,
\end{equation}
where the phase-drift coefficient is given by,
\begin{equation} \label{mu_phi}
\begin{aligned}
	\mu_{\phi}
=
	&\frac{1}{r}\sum_{n=-\infty}^{\infty}g_0A_L\left\{
	A_L  \operatorname{Re}
	\left[
		\frac{J_n J_{n-1}}{h_n h_{n-1}^*}
	\right]
\right.
\\
&
\left.
+
	A_e^{op}  \operatorname{Re}
	\left[
		\frac{e^{-i\phi}J_n J_{n-2}}{h_n h_{n-1}^*}e^{-i\epsilon t}
	\right]
+
	A_e^{op}  \operatorname{Re}
	\left[
		\frac{e^{i\phi}J_{n-1} J_{n-1}}{h_n h_{n-1}^*}e^{i\epsilon t}
	\right]
	\right\},
\end{aligned}
\end{equation}
and the explicit expressions for the radius-drift coefficient is given in Eq.~(\ref{radius_drift}). In the last expression $J_n := J_n(-2 g_0 r/\omega_m)$ is the Bessel function of the first kind of the $n$th order, $\epsilon\equiv \omega_e^{op} - \omega_m$ is the detuning between the frequencies of the reference drive and the natural frequency of the mechanical oscillator, and $h_n$ is defined as
\begin{equation} 
	h_n
=
	\frac{\gamma_c}{2}+i(n\omega_m-\Delta_\text{eff}),
\end{equation}
where the definition for the effective detuning of the cavity field, $\Delta_\text{eff}$, is given in Eq.~(\ref{effective_detuning_field}).

This EOM describes the dynamics of the mechanical oscillator and, in an appropriate parameter regime, will therefore describe the synchronization of the mechanical oscillator onto the optical reference drive. The onset of synchronization is characterized by the locking of the phase of the mechanical oscillator to the phase of the optical drive, while the radius of oscillation stays approximately constant. For that reason, we can neglect the term describing the drift of the radius, $\mu_r$, while focusing on the drift of the phase, Eq.~(\ref{mu_phi}). We are therefore left with 
\begin{equation}  \label{FPE_reduced}
	\dot{\sigma}
=
-
	\partial_\phi \mu_\phi
\sigma,
\end{equation}
from which we recognize that $\mu_\phi = \dot{\phi}$. Let us therefore focus on $\mu_\phi$, Eq.~(\ref{mu_phi}), which completely determines the time evolution of $\phi$. The first term is the known amplitude-dependent optomechanical frequency shift $\delta \omega$ (see Ref.~\cite{Armour}), i.e., we obtain
\begin{multline}
	\mu_{\phi}
=
	\dot{\phi}
=
	-\delta \omega
\\  \quad
+
	\frac{g_0A_LA_e^{op}}{r}\sum_n
	\operatorname{Re}
	\left[
		\frac{e^{-i(\phi+\epsilon t)}J_n J_{n-2}+e^{i(\phi+\epsilon t)}J_{n-1} J_{n-1}}{h_n h_{n-1}^*}
	\right]
	.
\end{multline}
In the sideband-resolved regime and with a detuning close to the mechanical frequency, i.e.,~$\gamma_c/2  \ll \Delta_\text{eff} \approx \omega_m$, terms with $h_1$ in the denominator are close to resonance. For that reason, we will keep only the terms with $n=1,2$. We then find
\begin{equation} \label{mu_phi_main}
	\dot{\phi}
=
	-\delta \omega
+
	A_e^{op,\text{eff}}
	\sin(\phi+\epsilon t),
\end{equation}
where we have shifted $\phi$ by a constant and defined the effective drive strength as
\begin{equation}
\begin{aligned}
	A_e^{op,\text{eff}}
&=
	\frac{g_0A_LA_e^{op}}{r\omega_m^2 \left(1+\frac{\gamma_c^2}{4\omega_m^2}\right)}\times
\\
& 
	\qquad \sqrt{(J_2+J_0)^2 J_0^2+\frac{4\omega_m^2}{\gamma_c^2}(J_2J_0 - 2J_1^2 - J_0^2)^2}
	.
\end{aligned}
\end{equation}
Adding the frequency difference $\epsilon$ to both sides of Eq.~(\ref{mu_phi_main}), we obtain the Adler equation
\begin{equation} \label{Adler_O_Main}
\begin{aligned}
	\dot{\delta \phi}
&=
	(\omega_e^{op}-\omega_m^\text{eff})
+
	A_e^{op,\text{eff}} \sin(\delta \phi),
\end{aligned}
\end{equation}
where the effective mechanical frequency is $\omega_m^\text{eff}\equiv \omega_m + \delta\omega$, and we have defined $\delta \phi \equiv \phi+\epsilon t$. Note that $\delta \phi = (\phi-\omega_mt) + \omega_e^{op}t$ is just the difference of phase of the mechanical oscillator (in a frame rotating with $\omega_m$) to the phase of the external drive.

\begin{figure}[t]
\includegraphics[width=\columnwidth]{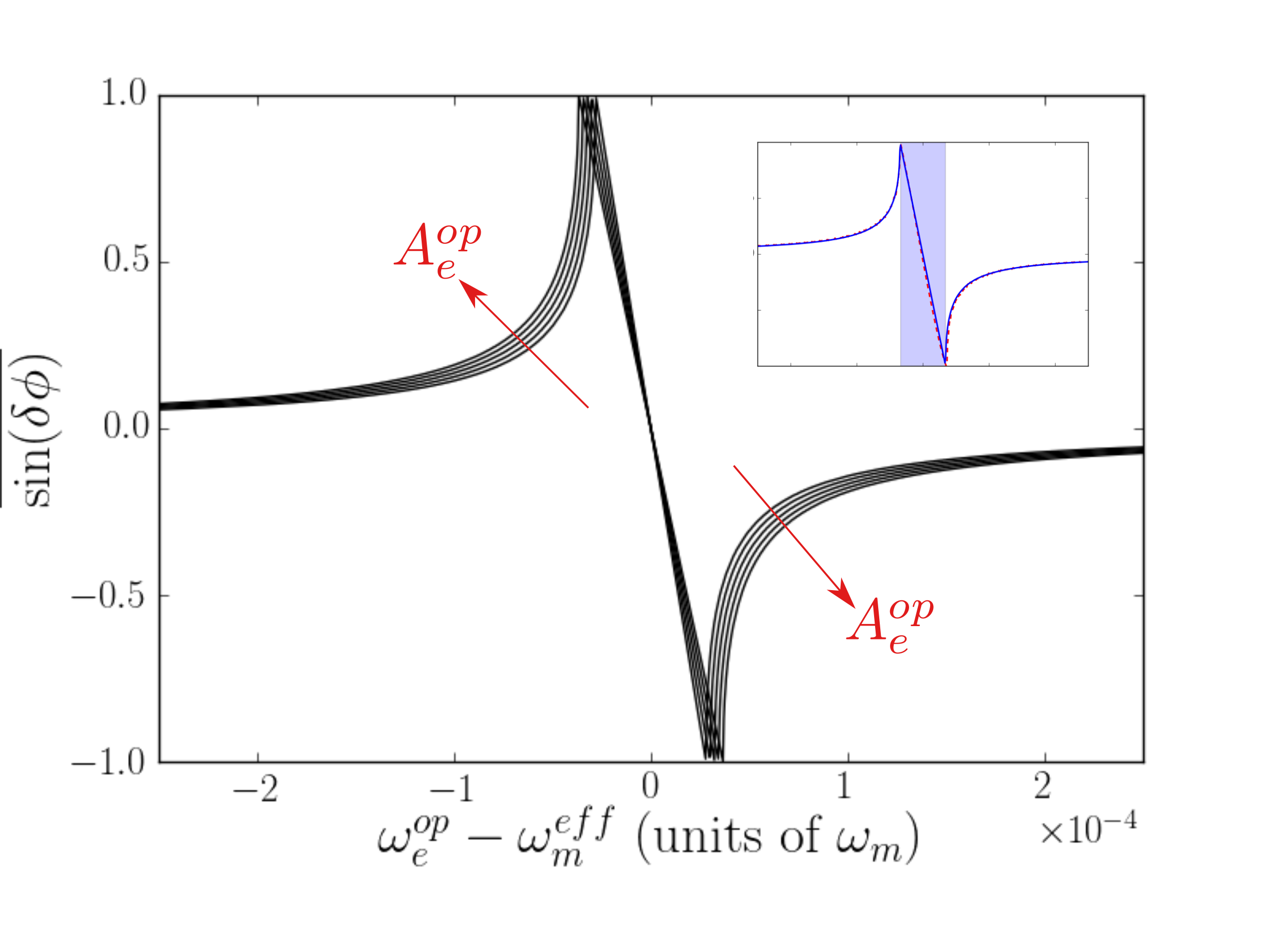}\caption{Synchronization of the mechanical self-oscillator to an optical reference drive. The analytically calculated time-average $\overline{\sin(\delta \phi)}$ as a function of $(\omega_e^{op}-\omega_m^\text{eff})$ for different values of $A_e^{op}$, from $0.13$ to $0.17$. The inset compares the analytical solution (blue) with the numerical simulation (red dashed) for $A_e^{op}=0.15$. It shows excellent agreement. Colored region indicates the synchronization region, $\mathrm{d}\delta \phi/\mathrm{d}t = 0$. The parameters of the optomechanical system are taken in the classical regime, $(g_0, \gamma_c, \gamma_m, A_L, \Delta, n_\text{th})=(0.015, 0.5, 0.0001, 1.0, 1.0, 0)\times \omega_m$.  
\label{Fig:Optical_Classical_Main}}
\end{figure}

The Adler equation describes the synchronization of the mechanical self-oscillator to the reference drive, as shown in Fig.~\ref{Fig:Optical_Classical_Main}, in which we plot $\overline{\sin{\delta\phi}}$ as a function of $(\omega_e^{op}-\omega_m^\text{eff})$ for different drive strengths, where the overline refers to time-averaging. For $|\omega_e^{op}-\omega_m^\text{eff}|<A_e^{op,\text{eff}}$, the solution to Eq.~(\ref{Adler_O_Main}) is $\dot{\delta \phi}=0$. Therefore phase-locking takes place. For $|\omega_e^{op}-\omega_m^\text{eff}|\gg A_e^{op,\text{eff}}$, $\sin(\delta\phi)$ time-averages to zero. The optomechanical parameters chosen in Fig.~\ref{Fig:Optical_Classical_Main} can be readily obtained in a wide range of experiments ~\cite{Kleckner, Chan, Aspelmeyer}. In Ref.~\cite{Kleckner} a mechanical resonator of frequency $\omega_m/(2\pi)=9.7(\text{kHz})$ was studied, while in Ref.~\cite{Chan} a mechanical resonator of frequency $\omega_m/(2\pi)=3.9(\text{GHz})$ was studied. In both, the parameters of the optomechanical system were similar to those given in Fig.~\ref{Fig:Optical_Classical_Main}.

We can further test this derived Adler equation by comparing it with the numerical prediction, which can be obtained by integrating the optomechanical equations of motion for the cavity field $\alpha$ and the mechanical field $\beta$ \cite{Armour}
\begin{align}
	\dot{\alpha}
&=
	i\Delta \alpha
+
	ig_0(\beta+\beta^*)\alpha
-
	\frac{\gamma_c}{2}\alpha
+
	A_L
+
	A_e^{op} e^{-i\omega_e^{op} t},
\\
	\dot{\beta}
&=
	ig_0|\alpha|^2
-
	i\omega_m \beta
-
	\frac{\gamma_m}{2}\beta.
\end{align}
The result is shown in the inset of Fig.~\ref{Fig:Optical_Classical_Main}. The synchronization region is indicated by the colored region. There is a very good agreement between the prediction of the derived microscopic equation and the numerical simulation. 

\begin{figure}[t]
\includegraphics[width=\columnwidth]{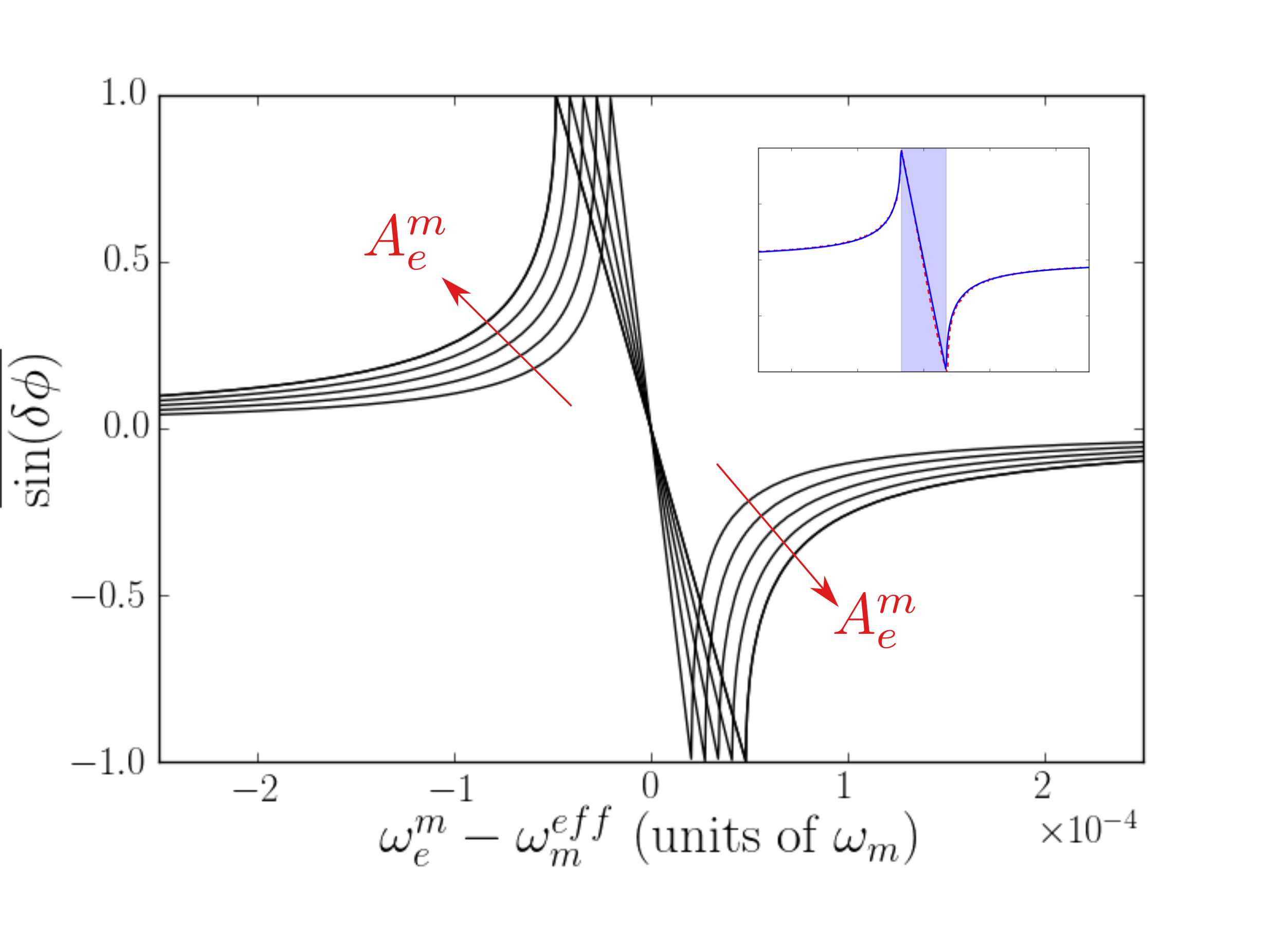}\caption{Synchronization of the mechanical self-oscillator to a mechanical reference drive. The analytically calculated time-average $\overline{\sin(\delta \phi)}$ as a function of $(\omega_e^{m}-\omega_m^\text{eff})$ for different values of $A_e^m$, from $0.003$ to $0.007$. The inset compares the analytical solution (blue) with the numerical simulation (red dashed) for $A_e^m=0.005$. It shows excellent agreement. Colored region indicates the synchronization region, $\mathrm{d}\delta \phi/\mathrm{d}t = 0$. The parameters of the optomechanical system are taken in the classical regime, $(g_0, \gamma_c, \gamma_m, A_L, \Delta, n_\text{th})=(0.01, 0.3, 0.0001, 1.0, 1.0, 0)\times \omega_m$.  
\label{Fig:r=145_main}}
\end{figure}

\textit{Case (2) - Mechanical drive} - Analogously to \emph{case (1)}, by applying the laser theory for optomechanical limit cycles we obtain an EOM for the phase distribution of the self-oscillator, $\sigma(r,\phi)$. This EOM has the same form as Eq.~(\ref{FP_equation_main}), but with a phase-drift coefficient which is given by
\begin{equation}
	\mu_{\phi}
=
	\frac{1}{r}
	\left\{
	\sum_n g_0A_L^2
	\operatorname{Re}
	\left[
		\frac{J_n J_{n-1}}{h_n h_{n-1}^*}
	\right]
	-
		A_e^m \sin \left[(\omega_e^m - \omega_m )t + \phi\right]
	\right\},
\end{equation}
and with a radius-drift coefficient which is given in the Appendix, Eq.~(\ref{mu_r_mech}). Now, taking identical steps to those shown in \emph{case (1)}, one reaches an Adler equation,
\begin{equation} \label{Adler_M}
\begin{aligned}
	\dot{\delta \phi}
&=
	(\omega_e^m - \omega_m^\text{eff})
+
	A_e^{m,\text{eff}} \sin(\delta \phi),
\end{aligned}
\end{equation}
where the effective drive strength is
\begin{equation}
	A_{e}^{m,\text{eff}}
=
	\frac{A_e^m}{r}.
\end{equation}
This form of the Adler equation agrees with \cite{Heinrich}. In Fig.~\ref{Fig:r=145_main} we plot $\overline{\sin{\delta\phi}}$ as a function of $(\omega_e^{op}-\omega_m^\text{eff})$ for different drive strengths, where the overline refers to time-averaging.
We can further test this analytical equation by comparing it with the classical numerical prediction, which can be obtained by integrating the equations of motion
\begin{align}
	\dot{\alpha}
&=
	i\Delta \alpha
+
	ig_0(\beta+\beta^*)\alpha
-
	\frac{\gamma_c}{2}\alpha
+
	A_L
,
\\
	\dot{\beta}
&=
	ig_0|\alpha|^2
-
	i\omega_m \beta
-
	\frac{\gamma_m}{2}\beta
+
	A_e^m e^{-i\omega_e^m t}.
\end{align}
The comparison is seen in the inset of Fig.~\ref{Fig:r=145_main}. A very good agreement is found between the analytical Adler equation and the numerical simulation.

\section{Quantum Synchronization - \\ Numerical Demonstration} \label{quant_sync}
The optomechanical system is theoretically suggested to demonstrate synchronization also in a quantum parameter regime, in which $g_0\ll \omega_m$ does not hold anymore. In that parameter regime, the quantum shot noise plays an important role, and cannot be neglected as in the previous section. The quantum synchronization of two such systems was theoretically studied in Ref.~\cite{Weiss}. Here we show numerically that the mechanical self-oscillator is expected to synchronize to a reference drive in the quantum parameter regime. Before discussing the numerical calculation and the results, we introduce the synchronization measure used.

\begin{figure}[t]
\includegraphics[width=\columnwidth]{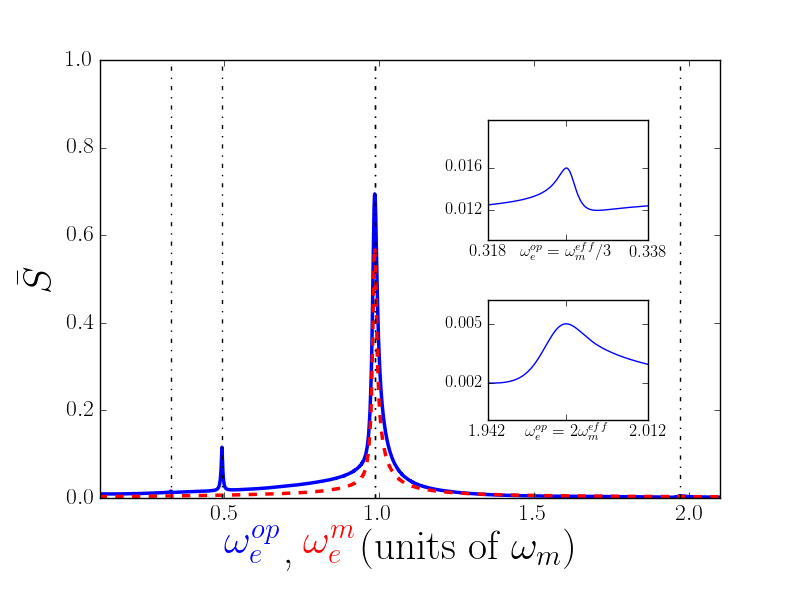}\caption{The time-averaged synchronization measure $\bar{S}$ as a function of the external drive's frequency, shown in blue for an optical drive with $A_e^{op}/\omega_m = 0.08$ and in a red dashed line for a mechanical drive with $A_e^m/\omega_m = 0.008$. For the mechanical drive there is only one synchronization peak at $\omega_e^m= \omega_m^\text{eff}$, while the optical drive leads to multiple synchronization peaks at $\omega_e^{op} = \left\{ \omega_m^\text{eff}/3, \omega_m^\text{eff}/2,  \omega_m^\text{eff},  2\omega_m^\text{eff}\right\}$. The black dotted lines are plotted at these frequencies. The synchronization peaks at $\omega_e^{op} = \left\{ \omega_m^\text{eff}/3,  2\omega_m^\text{eff}\right\}$ are hardly noticeable in the scale of the figure, and are therefore shown in the two insets. Optomechanical parameters are the same as in Fig.~\ref{Fig:Wigner_Bare}.
\label{Fig:Arnold_Tongue_Scan}}
\end{figure}
\subsection{The synchronization measure} \label{Our_synchronization_measure}
Synchronization of a self-oscillator to an external drive is the development of phase preference for the self-oscillator as it tends towards phase-locking to the phase of the external drive. As shown in Fig.~\ref{Fig:Wigner_Bare}, this phase preference is easily seen in the phase-space distribution of the mechanical oscillator. The information stored in the phase-space distribution can be accounted for by using a single number \cite{Loerch},
\begin{equation} \label{synchronization_measure}
	S
=
	\frac{|\braket{b}|}{\sqrt{\braket{b^\dagger b}}},
\end{equation}
where the bracket $\braket{\ldots}$ denotes averaging over the phase-space distribution.
The numerator holds information regarding the spread of the phase-space distribution, while the denominator is introduced for the purpose of normalization. For a completely phase-independent limit cycle centered around zero, the oscillator is obviously completely unsynchronized,  and indeed we will find $S=0$. As the self-oscillator synchronizes to an external drive, a phase preference develops. This will reduce the phase variation, resulting in larger values of $S$. For a coherent state the measure is $S=1$, meaning the oscillator is strongly synchronized to the drive. 
This measure cannot be used in cases for which the self-oscillator develops multiple preferred phases.

Note that in the optomechanical system, the self-oscillations developing in the mechanical oscillator are centered around some point in phase-space, $\beta_c$, which is generally different than the origin. This is seen in Fig.~\ref{Fig:Wigner_Bare}~(a). This deviation from the origin influences the synchronization measure, Eq.~(\ref{synchronization_measure}). This can be easily corrected and accounted for. To do so, we move to a displaced frame by using the displacement operator $D(-\beta_c)=\text{exp}(-\beta_c b^\dagger + \beta_c b)$. For the rest of this work, we will be working in the appropriate displaced frame. 

The problem of an optomechanical system with an additional reference drive, Eq.~(\ref{temp_master1}) with either Eq.~(\ref{optical_drive}) or Eq.~(\ref{mechanical_drive}), contains a time-dependent Hamiltonian. For that reason, a steady state does not emerge. However, in the late-time dynamics, the system evolves into a state which is periodic in time with periodicity $\tau\equiv2\pi/\omega_e^i$, where $i$ denotes the optical- or the mechanical-reference drive. This is true in the synchronized state and outside the synchronized state, and it is the result of the periodic time dependence of the Hamiltonian. For that reason, in the late-time dynamics the synchronization measure $S$ is a function of time with the same periodicity, $S(t)=S(t+\tau)$. The variation of $S$ over the time scale $\tau$ in the late-time dynamics is relatively small, and is of order $S\sim 0.01$ at maximum.
To conveniently discuss synchronization, we use $\bar{S}$, defined as the time-average of $S$ over a period $\tau$.

\begin{figure}[t]
\includegraphics[width=\columnwidth]{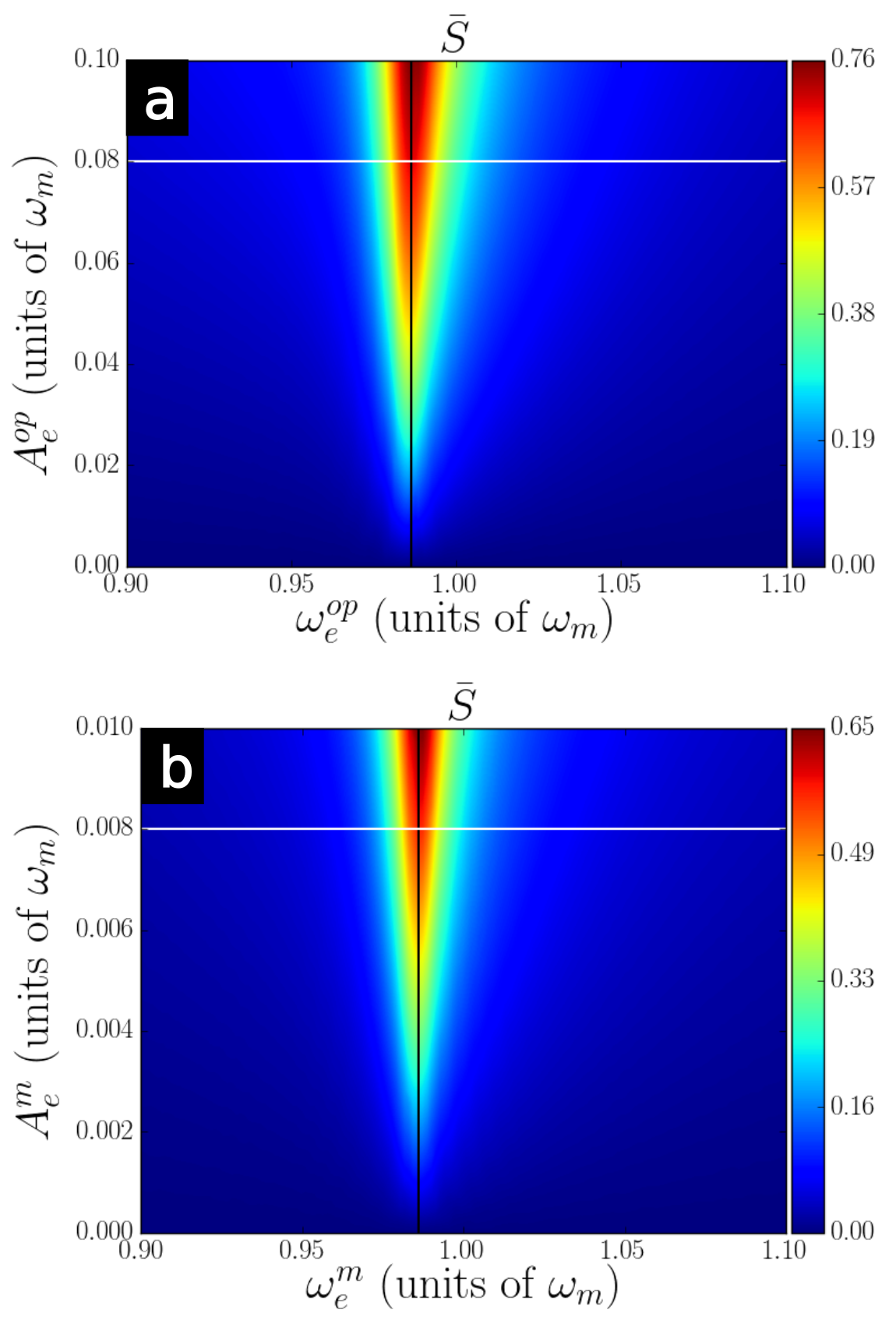}\caption{Arnold Tongue: The synchronization measure $\bar{S}$ is plotted as a function of the drive frequency and strength for case (1) (a), and for case (2) (b). $\bar{S}$ has the typical shape of an Arnold tongue. The black lines mark $\omega_e^{op} = \omega_m^\text{eff}$ and $\omega_e^m = \omega_m^\text{eff}$ in (a) and (b) correspondingly. The horizontal white lines mark the cut along which Fig.~\ref{Fig:Arnold_Tongue_Scan} is plotted. Optomechanical parameters are the same as in Fig.~\ref{Fig:Wigner_Bare}.
\label{Fig:Full_Arnold_Tongue}}
\end{figure}

\subsection{Numerical Results} \label{Sub:Numerical}

To numerically study synchronization of the mechanical self-oscillator to an external drive, we use QuTiP \cite{Johansson2, Johansson1}. 

\textbf{\textit{Case (1) - Optical laser drive}} - In Fig.~\ref{Fig:Arnold_Tongue_Scan}, the time-averaged synchronization measure $\bar{S}$ is plotted in blue as a function of the frequency of the reference drive, $\omega_e^{op}$. A main synchronization peak appears about an effective mechanical frequency, $\omega_m^\text{eff}$, slightly shifted from $\omega_m$. This shift of the mechanical frequency is known~\cite{Rodrigues, Armour} to be the result of the average dynamics of the electromagnetic cavity. Synchronization peaks at other frequencies are found as well: A synchronization peak about $\omega_e^{op} = \omega_m^\text{eff}/2$ is clearly visible, and in the insets of Fig.~\ref{Fig:Arnold_Tongue_Scan} we zoom in on the very small synchronization peaks at $\omega_e^{op} = \left\{ \omega_m^\text{eff}/3,  2\omega_m^\text{eff}\right\}$. These synchronization peaks are known in the literature as high-order synchronization~\cite{Pikovsky, Balanov}. While in principle high-order synchronization is always present when synchronizing a self-oscillator to a reference drive, it is in practice very difficult (if not impossible) to detect. The presence of a reference drive which contains many frequency components in its oscillation can enhance the synchronization peaks~\cite{Balanov}. As shown in the Appendix, the effective drive of the mechanical self-oscillator, Eq.~(\ref{drive_field}), indeed contains multiple frequencies. For that reason, and in contrast to case (2), we can observe the smaller synchronization peaks. We can also notice an asymmetry in the synchronization peak with respect to the reference field's frequency. This can be also be seen in Fig.~\ref{Fig:Full_Arnold_Tongue}. While there is no reason to expect perfect symmetry, it is visible that the case of an optical reference drive is more asymmetric. This is due to the high-order synchronization peaks.

In Fig.~\ref{Fig:Full_Arnold_Tongue}~(a) we focus on the synchronization peak for $\omega_e^{op} = \omega_m^\text{eff}$. This corresponds to the maximal synchronization peak shown in Fig.~\ref{Fig:Arnold_Tongue_Scan}.
The synchronization measure $\bar{S}$ is plotted as a function of both $A_e^{op}$ and $\omega_e^{op}$. Indeed, the ``Arnold tongue" is present, a signature for synchronization.

\textbf{\textit{Case (2) - Mechanical drive}} - The reference drive synchronizes the mechanical oscillator at frequency $\omega_e^m = \omega_m^\text{eff}$. This is shown by the red dashed curve in Fig.~\ref{Fig:Arnold_Tongue_Scan}. In contrast to the optical case, no high-order synchronization is seen. Indeed, as the mechanical drive is acting directly on the mechanical self-oscillator, its influence is harmonic. Therefore high-order synchronization is not detected~\cite{Pikovsky, Balanov}. 

In Fig.~\ref{Fig:Full_Arnold_Tongue}~(b) we focus on this synchronization peak. In this figure we vary both the external frequency $\omega_e^m$ and the strength of the external drive, $A_e^m$, and the ``Arnold tongue" is clearly observed.


\section{Conclusions}
In conclusion, our work fills a gap in the study of synchronization of an optomechanical system. Starting from the microscopic master equation, we analytically develop Adler equations describing the synchronization of the mechanical self-oscillator to a reference drive. This was done for two different reference drives, an optical one and a mechanical one (as was shown in Ref.~\cite{Heinrich}). We also show numerically that synchronization in a quantum parameter regime is expected, therefore suggesting the optomechanical system as a good candidate for the study of quantum synchronization.

\section*{Acknowledgments}

This work was financially supported by the Swiss SNF and the NCCR Quantum Science and Technology. A.N. holds a University Research Fellowship from the Royal Society and acknowledges support from the Winton Programme for the Physics of Sustainability.

\appendix

\section{Applying the Laser Theory for Optomechanical Limit Cycles} \label{App:Laser_Theory}

In the following, we apply laser theory for optomechanical limit cycles \cite{Loerch2} to our current problem, Eq.~(\ref{temp_master1}), with an additional reference drive as given by Eq.~(\ref{optical_drive}) or Eq.~(\ref{mechanical_drive}). We derive an EOM for $\sigma$, the phase-space distribution of the mechanical oscillator, for each case. In applying laser theory to our problem, the initial steps are identical to these taken when applying laser theory to a bare optomechanical system (with no reference drive). We therefore do not repeat these steps, but refer to Ref.~\cite{Loerch2} for our starting point, Eq.~(\ref{temp_master4}), which is presented below. A short summary of the steps taken in this Appendix:
\begin{itemize}
\item Switching to a phase-space representation for the mechanical oscillator degree of freedom. This allows us to use a different adiabatic reference state of the electromagnetic cavity field for each point in the phase-space of the mechanical oscillator. 
\item Assuming the electromagnetic cavity dynamics follows adiabatically the dynamics of the mechanical oscillator, we solve for an approximate solution for the cavity field. 
\item We use the solution for the cavity field as a reference state for the mechanical state, which allows us to obtain an EOM for $\sigma$.
\end{itemize}

\textit{Case (1) - Optical laser drive} - The master equation describing our system, Eq.~(\ref{temp_master1}) with an optical reference drive, Eq.~(\ref{optical_drive}), can be written in a phase-space picture for the mechanical oscillator \cite{Haake, Loerch2, Gardiner}. The system is then described by $\sigma(\beta, \beta^*)$, which is a density operator for the electromagnetic cavity field and a quasi-probability distribution for the mechanical oscillator, with $\beta$ representing the mechanical phase-space variable. This results in a dependence of the cavity detuning on the phase-space variables of the mechanical oscillator. If we further use the semipolaron transformation~\cite{Loerch2}, 
\begin{equation}
	\tilde{\sigma}(\beta, \beta^*, t)
=
	e^{g_0(\beta-\beta^*)a^\dagger a /\omega_m} \sigma(\beta, \beta^*, t)e^{-g_0(\beta-\beta^*)a^\dagger a /\omega_m}
,
\end{equation}
this dependence is conveniently transformed into one of the driving field.

The transformed master equation, in a mechanical frame rotating with frequency $\omega_m$, is then
\begin{equation}  \label{temp_master4}
	\partial_t \sigma(\beta,\beta^*,t)
=
	\left(
		L_m + L_c + L_\text{int}
	\right)
	\sigma(\beta,\beta^*,t),
\end{equation}
with
\begin{align} 
	L_m \sigma \label{temp_Lm}
&=
	\frac{\gamma_m}{2} \left(\partial_\beta \beta + \partial_{\beta^*}\beta^*\right)\sigma
+
	\gamma_m \left(n_\text{th} + 1/2\right) \partial^2_{\beta^* \beta}\sigma,
\\
	L_c \sigma
&=
	-i
	\left[
	-
		\Delta a^\dagger a
	-
		K(a^\dagger a)^2
	-
		i\left(E(t)^*a - E(t) a^\dagger\right)
	,
		\sigma
	\right]
\\ & \quad \nonumber
+
	\gamma_c D[a]\sigma,
\\ \label{temp_Lint}
	L_\text{int} \sigma
&=
	-i\frac{g_0}{2}
		\left(
			e^{i\omega_m t}
			\partial_\beta
		-
			e^{-i\omega_m t}
			 \partial_{\beta^*}
		\right)
		\sigma a^\dagger a
+
	H.c.
\end{align}
where the Kerr nonlinearity is $K=g_0^2/\omega_m$, and $E$ is a generalized driving field which depends on the mechanical state,
\begin{equation} \label{drive_field}
	E(t)
=
	\left(
		A_L
	+
		A_e^{op} e^{i\omega_e^{op} t}
	\right)
	\sum_{n=-\infty}^{\infty}
	J_n(-\eta r)
	e^{in(\omega_mt - \phi)}.
\end{equation}
Here, $J_n$ is the Bessel function of the first kind of the $n$th order, $r$ and $\phi$ are the mechanical phase-space variables $\beta=re^{i\phi}$, and $\eta = 2g_0/\omega_m$. We will use the shorthand notation $J_n := J_n(-\eta r)$.

We will now assume that the cavity dynamics, with a dominant time scale $\gamma_c^{-1}$, is fast compared to all other time scales in $L_m$ and $L_\text{int}$. This means that we can solve for the cavity field $\alpha(t)$, while assuming the state of the mechanical oscillator, described by the phase-space variables $\beta$ and $\beta^*$, is fixed. The solution for $\alpha(t)$ will then be considered as a classical reference amplitude. 

\subsection{Calculating the classical reference amplitude}

Under the assumption that the cavity dynamics with characteristic time scale $\gamma_c^{-1}$ is much faster than all other dynamics in the problem, its state is governed by $L_c$, while the effect of $L_\text{int}$ is neglected,
\begin{equation} \label{adiabatic}
	\dot{\sigma}
=
	L_c \sigma.
\end{equation}
This equation describes a cavity with Kerr nonlinearity, which is driven by two amplitude and phase modulated fields. An approximate solution to Eq.~(\ref{adiabatic}) can be found in two limits: the limit $|\alpha(t)|\gg 1$, i.e.,~a cavity which is driven by a strong drive to a state of large mean amplitude, and the limit $|\alpha(t)|\ll 1$, for which the cavity is driven by a weak drive and stays close to its ground state. In the former limit we neglect terms up to first order in $\alpha$, while in the latter we neglect terms of third order in $\alpha$:

\subsubsection{$|\alpha(t)|\gg 1$}
From Eq.~(\ref{adiabatic}), one can obtain an EOM for the classical cavity field amplitude $\alpha(t)$,
\begin{equation}\label{long_time_alpha_large}
	\dot{\alpha}(t) 
= 
	-\left\{\frac{\gamma_c}{2}-i\left[\Delta+2K|\alpha(t)|^2\right]\right\}\alpha(t)
+
	E(r, \phi, t).
\end{equation}
As a result of the form of the driving field $E$, Eq.~(\ref{drive_field}), we expect the solution to be of the form
\begin{equation} \label{alpha_solution}
	\alpha(r, \phi, t)
=
	\sum_{n=-\infty}^{\infty}
	\left[
		\alpha_n^l(r, \phi)e^{in\omega_mt}
	+
		\alpha_n^e(r, \phi)e^{i(n\omega_m+\omega_e^{op})t}
	\right],
\end{equation}
where the amplitudes $\alpha_n^l(r, \phi)$ and $\alpha_n^e(r, \phi)$ shall be determined.
As noticeable from Eq.~(\ref{long_time_alpha_large}), the effective detuning felt by the electromagnetic cavity depends on $|\alpha(t)|^2$. By placing the solution obtained for $\alpha(t)$, while keeping only the dominant dc components, we obtain for the effective detuning
\begin{equation} \label{effective_detuning_field}
	\Delta_\text{eff}(r, \phi)
=
	\Delta 
+ 
	2K\sum_n
	\left[
		|\alpha_n^l|^2+|\alpha_n^e|^2 +\alpha_n^l(\alpha_{n-1}^e)^*+(\alpha_{n+1}^l)^*\alpha_{n}^e
	\right],
\end{equation}
where we have assumed $\omega_e^{op}=\omega_m + \epsilon$, with $\epsilon \ll \omega_e^{op}, \omega_m$. This assumption is satisfied when studying synchronization in the vicinity of $\omega_e^{op}\approx\omega_m$.
We solve Eq.~(\ref{long_time_alpha_large}) by assuming a fixed effective detuning $\Delta_\text{eff}$. We then find,
\begin{align}
	\alpha_n^l   \label{alpha_n_l}
&=
	\frac{A_L}{h_n}J_n\left(-\eta r\right)e^{-in\phi},
\\
	\alpha_n^e
&=
	\frac{A_e^{op}}{h_{n+1}}J_n\left(-\eta r\right)e^{-in\phi},
\\				\label{h_n}
	h_n
&=
	\frac{\gamma_c}{2}+i(n\omega_m-\Delta_\text{eff}),
\end{align}
where we have used $\omega_e^{op}=\omega_m + \epsilon$ again. We have therefore found a solution for $\alpha(t)$.
\subsubsection{Displaced frame for $|\alpha(t)|\ll 1$}
In this limit, Eq.~(\ref{adiabatic}) dictates that $\alpha(t)$ should solve
\begin{equation}
	\dot{\alpha}(t)
=
	\left[
		i(\Delta+K)
	-
		\frac{\gamma_c}{2}
	\right]
	\alpha(t)
+
	E(r,\phi,t).
\end{equation}
As compared with Eq.~(\ref{long_time_alpha_large}), we see that a different effective detuning should be defined, $\Delta_K = \Delta + K$. Then, we can proceed as was done in the $|\alpha(t)|\gg 1$ limit. Results will be in complete analogy, and can be obtained by changing $\Delta_\text{eff} \rightarrow \Delta_K$.

\subsection{Obtaining the EOM}

After finding the solution for $\alpha(t)$, which will serve as a classical reference amplitude for the mechanical oscillator, our next goal is to obtain the EOM for the phase-space distribution of the mechanical oscillator, $\sigma(\beta, \beta^*)$. We notice that the dynamics of this phase-space distribution are governed by Eq.~(\ref{temp_Lm}) and Eq.~(\ref{temp_Lint}). 
By placing the solution for $\alpha(t)$ into Eq.~(\ref{temp_Lint}), one obtains an EOM for the phase-space distribution of the mechanical oscillator,
\begin{equation}
\begin{aligned} \label{W_app}
	\dot{\sigma}
&=
	ig_0\sum_n
	 \partial_{\beta^*}
	\left[
		\alpha_{n}^l (\alpha_{n-1}^l)^*
	+
		\alpha_{n}^l (\alpha_{n-2}^e)^* e^{-i\epsilon t}
	+
		\alpha_{n}^e (\alpha_{n}^l)^* e^{i\epsilon t}
	\right]
	\sigma
\\
&
	+ \text{H.c.},
\end{aligned}
\end{equation}
where we have neglected terms proportional to $\propto (A_e^{op})^2$, kept only dc terms, and have used $\omega_e^{op} = \omega_m + \epsilon$, where $\omega_e^{op}, \omega_m \gg \epsilon$. This allowed us to send $h_n^e \rightarrow h_{n+1}$, while keeping the exponentials depending on $\epsilon$, as they will be needed later to describe synchronization.

In describing limit cycles and synchronization, it is more natural to work in polar coordinates. We therefore transform Eq.~(\ref{W_app}) to a polar coordinate system. A more detailed description of this transformation can be found in Ref.~\cite{Loerch2}. The transformed EOM is then,
\begin{equation}  \label{FP_equaton}
	\dot{\sigma}
=
\left[
-
	\partial_r \mu_r
-
	\partial_\phi \mu_\phi
\right]
\sigma,
\end{equation}
where the drift coefficients are given by
\begin{align}
&\begin{aligned} \label{drift_phase}
	\mu_{\phi}
=
	&\frac{1}{r}\sum_ng_0A_L\left\{
	A_L  \operatorname{Re}
	\left[
		\frac{J_n J_{n-1}}{h_n h_{n-1}^*}
	\right]
\right.
\\&
\left.
+
	A_e^{op}  \operatorname{Re}
	\left[
		\frac{e^{-i\phi}J_n J_{n-2}}{h_n h_{n-1}^*}e^{-i\epsilon t}
	\right]
+
	A_e^{op}  \operatorname{Re}
	\left[
		\frac{e^{i\phi}J_{n-1} J_{n-1}}{h_n h_{n-1}^*}e^{i\epsilon t}
	\right]
	\right\},
\end{aligned}
\\
&\begin{aligned}\label{radius_drift}
	\mu_{r}
=
-&
	\frac{\gamma_m}{2} r
+
	\sum_ng_0A_L\left\{
	A_L  \operatorname{Im}
	\left[
		\frac{J_n J_{n-1}}{h_n h_{n-1}^*}
	\right]
\right.
\\&
\left.
+
	A_e^{op}  \operatorname{Im}
	\left[
		\frac{e^{-i\phi}J_n J_{n-2}}{h_n h_{n-1}^*}e^{-i\epsilon t}
	\right]
+
	A_e^{op} \operatorname{Im}
	\left[
		\frac{e^{i\phi}J_{n-1} J_{n-1}}{h_n h_{n-1}^*}e^{i\epsilon t}
	\right]
	\right\},
\end{aligned} 
\end{align}
where we have neglected terms $\propto 1/r$ in the equation for $\mu_r$ and terms $\propto 1/r^2$ in the equation for $\mu_\phi$, and have included the effect due to Eq.~(\ref{temp_Lm}). In the limit of $A_e^{op}\rightarrow 0$, one retrieves the known expression from \cite{Loerch2}.

\textit{Case (2) - Mechanical drive} - In applying the laser theory for optomechanical limit cycles for this case, we take steps completely analogous to those taken in the previous case. As the mechanical reference drive acts directly on the mechanical self-oscillator, it does not appear in the solution for $\alpha(t)$ nor in the elimination of the electromagnetic cavity. This fact makes calculations more straightforward in the present case, and we do not explicitly present them here. The EOM obtained has the same form as Eq.~(\ref{FP_equaton}), with drift coefficients which are given by
\begin{align}
&\begin{aligned}  \label{mu_phi_m}
	\mu_{\phi}
&=
	\frac{1}{r}
	\left\{
	\sum_n g_0A_L^2
	\operatorname{Re}
	\left[
		\frac{J_n J_{n-1}}{h_n h_{n-1}^*}
	\right]
	-
		A_e^m \sin \left[(\omega_e^m - \omega_m )t + \phi\right]
	\right\},
\end{aligned}
\\
&\begin{aligned} \label{mu_r_mech}
	\mu_{r}
&=
-
	\frac{\gamma_m}{2} r
+
	\sum_ng_0A_L^2
	\operatorname{Im}
	\left[
		\frac{J_n J_{n-1}}{h_n h_{n-1}^*}
	\right]
+
	A_e^m \cos \left[(\omega_e^m - \omega_m )t + \phi\right].
\end{aligned}
\end{align}
As in the previous case, we have neglected terms $\propto 1/r$ in the equation for $\mu_r$ and terms $\propto 1/r^2$ in the equation for $\mu_\phi$.
In the limit of $A_e^m\rightarrow 0$, one retrieves the known expressions from \cite{Loerch2}.

\section{Fokker-Planck Equation for the Mechanical Self-Oscillator}
Using laser theory for optomechanical systems allows one to obtain a Fokker-Planck equation (FPE) describing the dynamics of the mechanical self-oscillator. This FPE is of the form
\begin{equation}
	\dot{W}
=
\left[
-
	\partial_r \mu_r
-
	\partial_\phi \mu_\phi
+
	\partial^2_{rr} D_{rr}
+
	\partial^2_{r\phi} D_{r\phi}
+
	\partial^2_{\phi \phi} D_{\phi\phi}
\right]
W,
\end{equation}
where $W(r, \phi)$ is chosen to be the Wigner phase-space distribution, $\mu_r$ and $\mu_\phi$ are the drift coefficients of the phase-space variables $r$ and $\phi$ correspondingly, and $D_{rr}$, $D_{r\phi}$ and $D_{\phi\phi}$ are the diffusion coefficients.
In App.~\ref{App:Laser_Theory} we aimed to obtain only the drift coefficients, as they are sufficient to describe synchronization in a parameter regime in which the diffusion does not play a significant role. For completion and for those interested, we give in this appendix the expressions for the diffusion coefficients.

\textit{Case (1) - Optical laser drive} - The drift coefficients of the FPE equation are given in Eqs.~(\ref{drift_phase})-(\ref{radius_drift}), while the diffusion coefficients are given by
\begin{widetext}
\begin{align}
&\begin{aligned} \label{Diff_Phase}
	D_{\phi\phi}
&=
	\frac{1}{r^2}\frac{\gamma_m(n_\text{th}+1/2)}{4}	
+
	\frac{1}{r^2}\sum_n\frac{\gamma_c g_0^2 A_L}{8|\tilde{h}_{n+1}|^2}
	\left\{
		A_L \frac{J_{n}^2}{|h_{n}|^2}
	+
		A_L \frac{J_{n+2}^2}{|h_{n+2}|^2}
	+
		2A_L\operatorname{Re}
		\left[
			\frac{J_{n+2} J_n }{h_n h_{n+2}^*}
		\right]
	+
		2A_e^{op}\frac{J_n}{|h_{n}|^2}
		\left(
			J_{n+1}
		+
			J_{n-1}
		\right)
		\cos(\phi + \epsilon t)
	\right.
	\\&
	\left.
	\qquad\qquad\qquad\qquad\qquad\qquad\qquad\qquad\qquad\qquad+
		2A_e^{op}\operatorname{Re}
		\left[
			\frac{e^{i\phi}J_{n+2} J_{n-1} }{h_n h_{n+2}^*}e^{i\epsilon t}
		\right]
	+
		2A_e^{op}\operatorname{Re}
		\left[
			\frac{e^{-3i\phi}J_{n} J_{n+1} }{h_n h_{n+2}^*}e^{-i\epsilon t}
		\right]
	\right\},
\end{aligned}
\\
&\begin{aligned}
	D_{r\phi}
&=
	-\frac{1}{r}\sum_n\frac{\gamma_c g_0^2 A_L}{2|\tilde{h}_{n+1}|^2}
	\left\{
		A_L\operatorname{Im}
		\left[
			\frac{J_{n+2} J_n }{h_n h_{n+2}^*}
		\right]
	+
		A_e^{op}\operatorname{Im}
		\left[
			\frac{e^{i\phi}J_{n+2} J_{n-1} }{h_n h_{n+2}^*}e^{i\epsilon t}
		\right]
	+
		A_e^{op}\operatorname{Im}
		\left[
			\frac{e^{-3i\phi}J_{n} J_{n+1} }{h_n h_{n+2}^*}e^{-i\epsilon t}
		\right]
	\right\},
\end{aligned}
\\
&\begin{aligned} \label{D_rr}
	D_{rr}
&=
	\frac{\gamma_m(n_\text{th}+1/2)}{4}
	+
	\sum_n\frac{\gamma_c g_0^2 A_L}{8|\tilde{h}_{n+1}|^2}
	\left\{
		A_L \frac{J_{n}^2}{|h_{n}|^2}
	+
		A_L \frac{J_{n+2}^2}{|h_{n+2}|^2}
	-
		2A_L\operatorname{Re}
		\left[
			\frac{J_{n+2} J_n }{h_n h_{n+2}^*}
		\right]
	+
		\frac{2A_e^{op} J_n}{|h_{n}|^2}
		\left(
			J_{n+1}
		+
			J_{n-1}
		\right)
		\cos(\epsilon + \delta t)
	\right.
	\\&
	\left.\qquad\qquad\qquad\qquad\qquad\qquad\qquad\qquad\qquad\qquad
	-2
		A_e^{op}\operatorname{Re}
		\left[
			\frac{e^{i\phi}J_{n+2} J_{n-1} }{h_n h_{n+2}^*}
			e^{i\epsilon t}
		\right]
	-2
		A_e^{op}\operatorname{Re}
		\left[
			\frac{e^{-3i\phi}J_{n} J_{n+1} }{h_n h_{n+2}^*}
			e^{-i\epsilon t}
		\right]
	\right\}.
\end{aligned}
\end{align}
\end{widetext}
\textit{Case (2) - Mechanical drive} - The drift coefficients of the FPE equation are given in Eqs.~(\ref{mu_phi_m})-(\ref{mu_r_mech}), while the reference field $A_e^m$ does not enter the expressions for the diffusion. The diffusion coefficients are therefore given in Eqs.~(\ref{Diff_Phase})-(\ref{D_rr}), with $A_e^{op}=0$.

\bibliographystyle{apsrev}
 \bibliography{Optomechanical_Sync_Drive}

\end{document}